\documentclass{elsart}
\usepackage{epsfig}
\usepackage{graphicx}
\usepackage{amssymb}
\usepackage{bm}

\begin{document}

\begin{frontmatter}

\title{D meson production in d+Au process using a perturbative approach}

\author[ifufrgs]{M. B. Gay Ducati \thanksref{gay}},
\author[ifmufpel]{V. P. Gon\c calves \thanksref{victor}},
\author[ifufrgs]{L. F. Mackedanz \thanksref{thunder}}

\thanks[gay]{E-mail:gay@if.ufrgs.br}
\thanks[victor]{E-mail:barros@ufpel.edu.br}
\thanks[thunder]{E-mail:thunder@if.ufrgs.br}

\address[ifufrgs]{Instituto de F\'{\i}sica, Universidade Federal do Rio Grande do Sul,
Caixa Postal 15051, CEP 91501-970, Porto Alegre, RS, Brazil}
\address[ifmufpel]{Instituto de F\'{\i}sica e Matem\'atica,  Universidade
Federal de Pelotas, Caixa Postal 354, CEP 96010-090, Pelotas, RS, Brazil}

\begin{abstract}
The D meson production at forward rapidities in d+Au processes is calculated using a pQCD
based model, assuming that this treatment could be used as a baseline for distinct dynamical
and medium effects. It is analysed how the nuclear effects in the nuclear partonic distributions
may affect this process at RHIC and LHC energies. An enhancement in the moderate $q_T$ region for RHIC,
due to anti-shadowing in the nuclear medium, is found. Our prediction for LHC suggests that
shadowing will suppress the D meson spectra for $q_T<14$ GeV.
\end{abstract}

\end{frontmatter}

Relativistic heavy ion collisions provide an opportunity to study
the QCD properties at energy  densities several hundred times
higher than the density of the atomic nuclei \cite{rafelski,wong}.
In these extreme conditions a deconfined state of quarks and
gluons, the Quark Gluon Plasma (QGP), is expected to be formed in
the early stage of the collision. These higher densities could
induce a large amount of energy loss by gluon bremsstrahlung while
hard partons propagate through the medium. The energy loss
experienced by a fast parton may serve as a measure of the density
of color charges of the medium it travels through
\cite{guywang,bdmps}. In a dense medium, as the QGP, the energy
loss may be huge. As large transverse momentum partons are
produced very early in these processes, one expects that they can
probe the early stage of the formed dense medium \cite{fragility}.
The mechanism of energy loss is thought to explain the observed
suppression of the high transverse momenta ($q_T$) hadron spectra
in central Au+Au collisions at RHIC \cite{expAAdata}. However,
this feature of data could be explained \cite{klm} as well by
saturation effects in the initial nuclear wavefunctions using the
Color Glass Condensate (CGC) formalism \cite{CGCreview}.

In order to determine which mechanism is responsible for this
suppression, d+Au collisions were studied at RHIC. At mid-rapidity
the data \cite{phenix3,star3,brahms2} show an absence of jet
quenching, which indicates that the observed high-$q_T$
suppression patterns in Au+Au collisions are not initial state
effects encoded in the wavefunction of the beam nucleus, but are
caused by final state interaction of hard partons with the
produced dense medium. In order to test the consistency of the
interpretation of these quenching effects due to energy loss in a
deconfined medium, a comparative study of the attenuation patterns
for massless and heavy partons was proposed
\cite{dokskhar,wang2zhang,daineseEPJC}. The large mass of heavy
quarks modifies the gluon bremsstrahlung, since it is suppressed
for small angles $\theta<m_Q/E$ \cite{deadcone}, where $m_Q$ is
the mass of heavy quark and $E$, its energy, implying different
energy losses for heavy and light quarks propagating in a dense
medium. As a consequence, one can observe a softening in the light
hadron spectra accompanying heavy quark jets, and a hardening of
the leading charmed hadrons \cite{dokskhar,wang2zhang,heavyloss}.
The production of heavy mesons are also affected by initial state
effects, and their magnitude has to be estimated for a realistic
prediction of energy loss of heavy quarks.

In this  work, the validity of the perturbative QCD and the
collinear factorization is assumed for RHIC kinematical regime,
and this treatment is considered as a baseline to explicitate the
presence of new dynamical effects in charmed meson production in
relativistic heavy ion collisions. Focus is given to hadron
(deuteron) - nucleus processes, since studies at this kind of
interactions can provide important benchmarks for further
measurements in nucleus-nucleus processes. In particular, the
forward rapidity region is studied, where the nuclear parton
momentum fraction, $x_2$, reaches smaller values and saturation
effects are expected to became important \cite{cronin,jamal,kkt2}.
Since the $x$ values reached, at RHIC energies, are not very
small, it is necessary to keep in mind what are the predictions of
the conventional QCD models which assume nuclear shadowing. In
what follows our analysis concerns to the rapidity dependence of
the nuclear modification ratios, $R_{\mathrm{AB}}$, defined by
\vspace*{-0.3cm}
\begin{eqnarray}\label{rabratio}
R_{\mathrm{AB}}(q_T)=\frac{d\sigma_{\mathrm{AB}}/dy
d^2q_T}{ABd\sigma_{\mathrm{pp}}/dy d^2q_T}\,\,,
\end{eqnarray}
where $y$ is the rapidity and $A$ and $B$ are atomic mass numbers,
considering the EKS parameterization \cite{eks} of nuclear
effects. Predictions for rapidity distributions for D mesons at
RHIC ($\sqrt{s}=200$ GeV) are also calculated and the analysis is
extended for LHC ($\sqrt{s}=5.5$ TeV). A comment related to the
light hadron production is in order here. Currently, the
description of the experimental results in the central rapidity
region can be obtained  using a perturbative approach which
includes  the nuclear shadowing effects in the partonic
distributions and an intrinsic transverse momentum of the
colliding partons  in order to reproduce  the Cronin peak 
\cite{vitev,accardiPLB,vogtratio}.
However, it is important to emphasize that the existing
conventional nuclear shadowing models cannot completely explain
both Cronin effect and the suppression in the forward region, as
well as the large value of that suppression. In contrast, in the
framework of the Color Glass Condensate \cite{CGCreview} both
effects are predicted \cite{cronin} to follow from the nonlinear
evolution equation. If a similar scenario will be present in
charmed hadron production is a subject of intense study
\cite{kharzeevNPA,tuchin,gvheavy}, since the bulk of the heavy
quark cross section comes from larger  $x$ values than for light
hadron production in all rapidity range \cite{pAhardprobes}.

Hadrons with heavy quarks are a very important tool to study the
properties of the strong interactions. Their large quark masses
provide a scale which allows the use of perturbative QCD for
computing production processes, since the long distance dynamics
is effectively decoupled from the short distance dynamics. The
value of the charm quark mass is in the limit of applicability of
perturbative QCD, being a matter of discussion. In this framework,
one can use the collinear factorization to calculate the heavy
quark production. In leading twist, the semi-inclusive cross
section factorizes into the product of gluon distributions, heavy
quark fragmentation function and the hard partonic cross section.
The application of factorization, however, is not evident for the
kinematic regime where the heavy quark mass $m_Q$ is much smaller
than the center of mass energy, $\sqrt{s}$ \cite{gvheavy}. For
instance, it has been claimed that in RHIC energy range, the $x$
values reached by the nucleons are lower enough to justify the
calculation with the nucleus assumed as a saturated dense partonic
system (the CGC), with a characteristic saturation scale $Q_s$,
breaking the factorization of the process \cite{kharzeevNPA}. In
particular, the heavy quark production, and consequently the D
mesons production, was studied using this semi-classical approach
in the Ref. \cite{kharzeevNPA,tuchin}, showing that the saturation
phenomenon makes the spectra harder as compared to the PYTHIA
prediction \cite{kharzeevNPA}.

For jet production in a hadronic collision a pQCD-based model \cite{accardiPLB} is used in this work.
In leading order pQCD, the $pp'$ inclusive cross section (where $p$ and $p'$ stand for a  proton ($p$) or a
nucleon ($N$)) for production of a parton of flavour $i = g,q,\bar{q}$ ($q=u,d,s,\dots$) with transverse
momentum $p_T$ and rapidity $y$ \cite{EskHonNPA} is written as a sum of contributions of the cross sections
coming from projectile ($p$) partons and from target ($p'$) partons:
\vspace*{-0.6cm}
\begin{eqnarray}\label{ppcoll_pt}
 \hspace*{-0.8cm} \frac{d\sigma}{dp_T^2 dy}^{\hspace{-0.3cm}pp'\rightarrow iX}
    \hspace{-0.7cm} &=&
    \langle{xf_{i/p}}\rangle_{y_i,p_T} \, \frac{d\sigma^{\,ip'}}{dy_i d^2p_T}\Big|_{y_i=y}
  + \langle{xf_{i/p'}}\rangle_{y_i,p_T} \,
    \frac{d\sigma^{\,ip}}{dy_i d^2p_T}\Big|_{y_i=-y} \ ,
\end{eqnarray}
where
\vspace*{-0.6cm}
\begin{eqnarray}\nonumber
\hspace*{-0.8cm} \langle{xf_{i/p}}\rangle_{y_i,p_T} &=& \frac{K}{\pi}
      \sum_j \frac{1}{1+\delta_{ij}}
      \int dy_2 x_1f_{i/p}(x_1,Q_p^2)
      \frac{d\hat\sigma}{d\hat t}^{ij} \hspace{-0.2cm}
      (\hat s,\hat t,\hat u)\\\label{avflux}
 &\times & x_2f_{j/p'}(x_2,Q_p^2)
    \Bigg/
      \frac{d\sigma^{ip'}}{d^2p_T dy_i}
    \\*
\hspace*{-0.8cm} \frac{d\sigma^{ip'}}{d^2p_T dy_i} &=& \frac{K}{\pi}
    \sum_j \frac{1}{1+\delta_{ij}}
    \int dy_2 \frac{d\hat\sigma}{d\hat t}^{ij} \hspace{-0.2cm}
    (\hat s,\hat t,\hat u) \, x_2f_{j/p'}(x_2,Q_p^2)
 \label{iNxsec}
\end{eqnarray}
are interpreted, respectively, as the average flux of incoming partons of
flavour $i$ from the hadron $p$, and the cross section for the parton-hadron
scattering. The rapidities of the $i$ and $j$ partons in the final state
are labelled by $y_i$ and $y_2$. In this model infrared regularization is performed by
adding a small mass to the gluon propagator and defining $m_{T}=\sqrt{p_T^2+p_0^2}$.
The fractional momenta of the colliding partons $i$ and $j$ are
$x_{1,2}=\frac{m_T}{\sqrt s}({\rm e}^{\pm y_i}+{\rm e}^{\pm y_2})$,
with the integration region for $y_2$ given by $-\log(\sqrt s/m_T-{\rm e}^{-y_i})\le
y_2\le \log(\sqrt s/m_T-{\rm e}^{y_i})$. More details are given in Refs. \cite{accardiPLB,EskHonNPA}.

For the charmed meson production at high energies the dominant subprocess is
$gg \rightarrow c\bar{c}$. This cross section $d\hat\sigma^{ij}/d\hat t$ can
be found, e.g., in \cite{combridge} and is proportional to $\alpha_{\rm s}
(\mu^2)$, with $\mu = Q_p = \sqrt{m_T^2+m_Q^2}$. The factor $K$ in (\ref{ppcoll_pt}) is introduced in
order to account for next-to-leading order (NLO) corrections and is, in general,
energy and scale dependent \cite{EskHonNPA}. For the parton distributions the
CTEQ5 parameterization at leading order \cite{CTEQ5}, evaluated at $Q_p$,
will be used and when the nuclear shadowing effects are considered in
the calculation, the EKS parameterization \cite{eks} will be employed.

Inclusive hadron production through independent fragmentation of the parton
$i$ into a hadron $h$, is computed as a convolution of the partonic cross
section (\ref{ppcoll_pt}) with a fragmentation function $D_{i\rightarrow h}(z,Q_h^2)$:
\vspace*{-0.6cm}
\begin{eqnarray} \label{pptohadron}
  \frac{d\sigma^{pp'\rightarrow hX}}{dq_T^2dy_h}
     = 
    \frac{d\sigma^{pp'\rightarrow iX}}{dp_T^2 dy_i}
    \otimes D_{i\rightarrow h}(z,Q_h^2) \ ,
\end{eqnarray}
where $q_T$ is  the transverse momentum of the hadron $h$, $y_h$ its rapidity,
and $z$ the light-cone fractional momentum of the hadron and of its parent
parton $i$. For details, see Eqs.~(8)-(11) of \cite{EskHonNPA}. This pQCD-based
model has been successful in describing the data for charged hadrons and neutral
pions, at mid-rapidity \cite{accardiPLB}. In the low $q_T$ region, it was considered an intrinsic
$k_T$ for the colliding partons, in order to correct the curvature of the hadron
spectrum. However, since the interest here is the modifications due to nuclear
shadowing, the intrinsic $k_T$ is not considered in this calculation. For the
fragmentation function, the Peterson function \cite{peterson} will be used with $\epsilon = 0.043$,
as in Ref. \cite{kharzeevNPA} in the CGC framework.

Since in heavy quark production at high energies the dominant
process is the gluon fusion, the cross section is strongly
dependent of the behavior of the nuclear gluon distribution.
Currently, there are several parameterizations in the literature
which predict distinct behaviors and magnitude of the nuclear
effects in  the gluon distribution and a recent comparison is
given in Ref. \cite{pAhardprobes}. For example, the EKS
parameterization \cite{eks} has a strong anti-shadowing ($R_g^A
\equiv xG_A/AxG_N >1$) at intermediate $x$ ($x \sim 0.1-0.2$), due
to momentum conservation constraint, and the EMC effect ($R_g^A
<1$) at $x \sim 0.2-0.8$. For lower values of $x$, it presents
shadowing ($R_g^A <1$). On the other hand, the HKM one \cite{hkm},
presents less shadowing at small $x$ values and the EMC effect is
not present at intermediate $x$. Furthermore, the momentum sum
rule is underestimated by the HIJING parameterization
\cite{hijing}, due to a strong gluon shadowing and a lack of
anti-shadowing effect (For a recent NLO analysis see Ref.
\cite{florian}). Due to these differences between the
parameterizations, only bounds can be estimated for nuclear
effects and the EKS one is used in order to provide a conservative
estimate. The distinct effects in different $x$ regions present in
this parameterization create an asymmetry in the rapidity
distribution: at large negative rapidities, the nuclear momentum
fraction, $x_2$, is large, while $x_1$ is small; conversely, the
positive rapidities access small $x_2$ and large $x_1$. Since our
goal is to study the nuclear modifications, our analysis deals
with positive rapidities. A similar study for light hadrons is
presented in Ref. \cite{wangrapda,vogtratio} (For an interesting
discussion about this subject see Ref. \cite{guzey}).

\begin{figure}
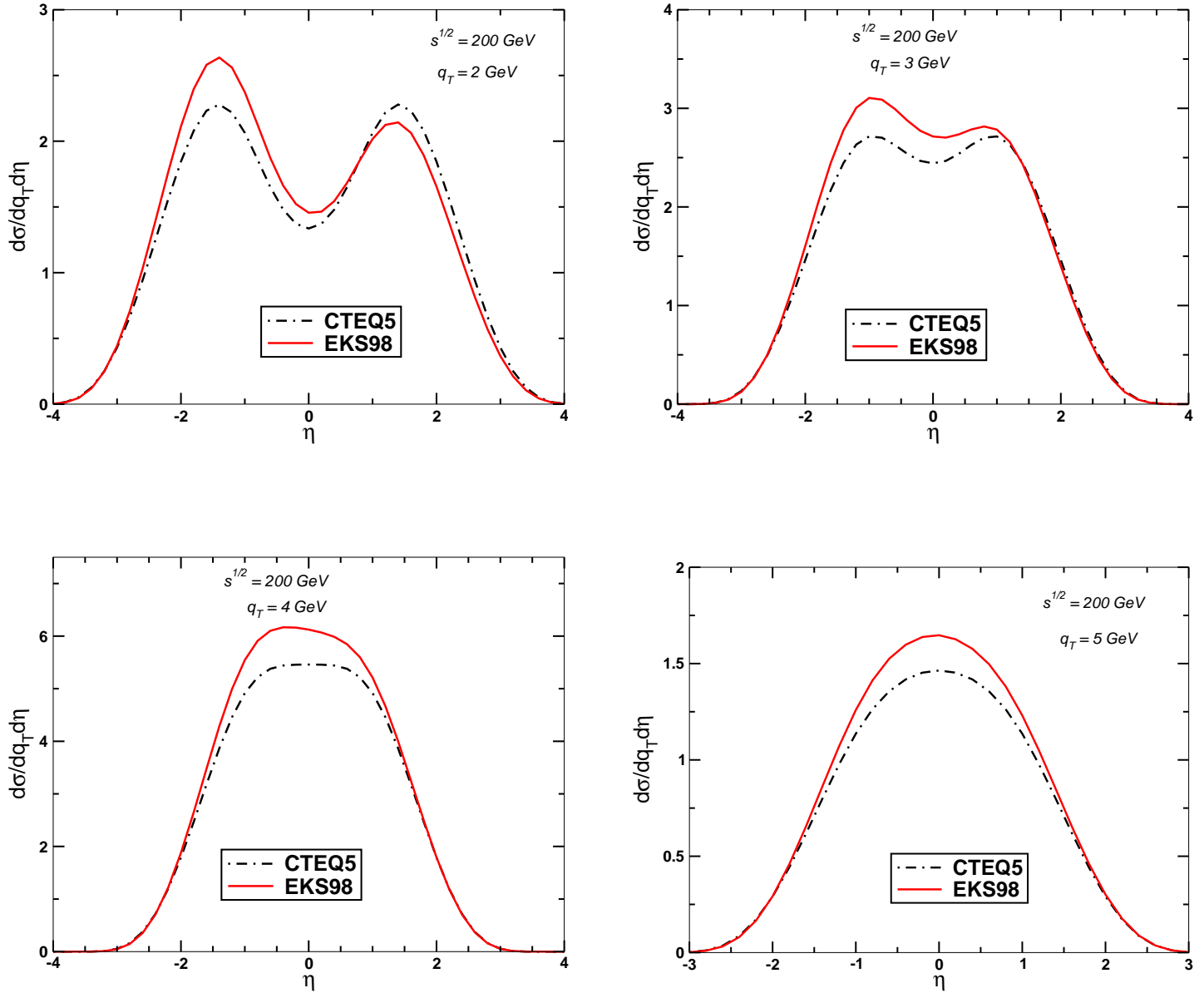

\centerline{
\begin{tabular}{ccc}
{\psfig{file=dscdyRHICpt2.eps,width=90mm}} & \,\, &
{\psfig{file=dscdyRHICpt3.eps,width=90mm}}\\
 & & \\
 & & \\
{\psfig{file=dscdyRHICpt4.eps,width=90mm}} & \,\, &
{\psfig{file=dscdyRHICpt5.eps,width=90mm}}
\end{tabular}}
\caption{Rapidity distributions for RHIC ($\sqrt{s}=200$ GeV) for distinct values of the transverse momentum.}
\label{fig0}
\end{figure}

In Fig. \ref{fig0} we present the rapidity distributions of the D meson
spectra at four distinct values of $q_T$ for d+Au collisions at $\sqrt{s}=200$
GeV. The dot-dashed curve, labelled CTEQ5, shows the prediction without nuclear
shadowing and the solid curve, labelled EKS98, shows the predictions
when the nuclear shadowing is considered. An asymmetry is observed at
low $q_T$, but disappears for higher $q_T$. At $q_T=2$ GeV, a strong
anti-shadowing enhances the spectra at negative rapidities, and shadowing
suppresses it for positive ones. For increasing $q_T$, this asymmetry is
weakened, with the $x_2$ values at positive rapidities increasing, entering
in the anti-shadowing region. At $q_T=5$ GeV, the rapidity symmetry is recovered.
Higher values of $q_T$ presents the reversion in the rapidity asymmetry, with
enhancement of spectra at positive rapidities.

\begin{figure}
\centerline{
\begin{tabular}{ccc}
{\psfig{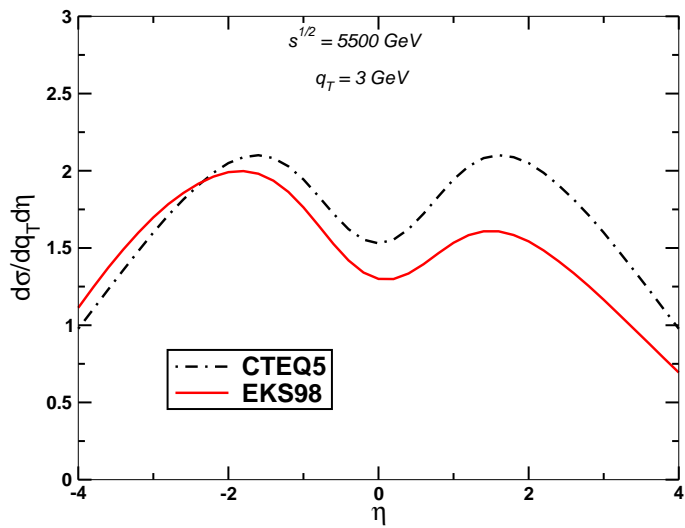}} & \,\, &
{\psfig{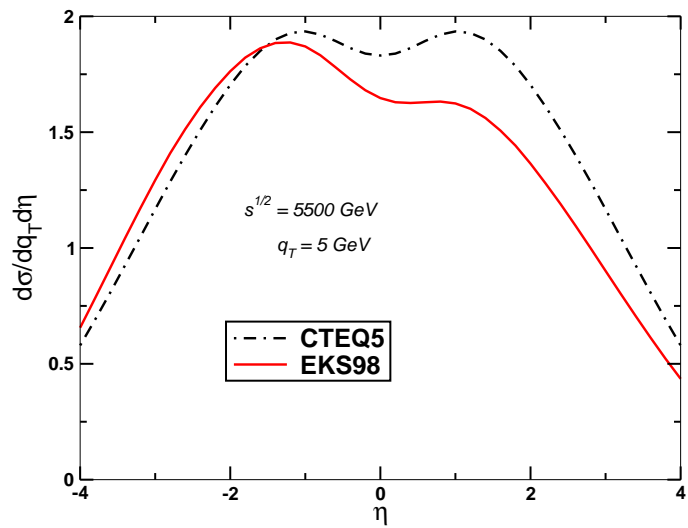}}\\
& & \\
& & \\
{\psfig{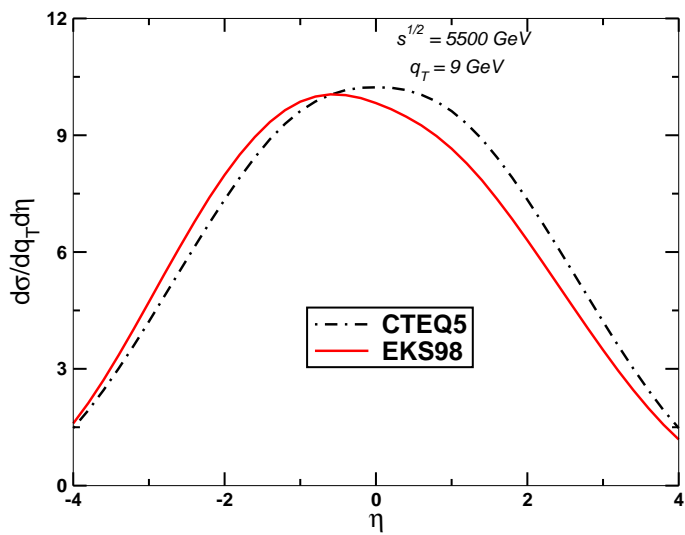}} & \,\, &
{\psfig{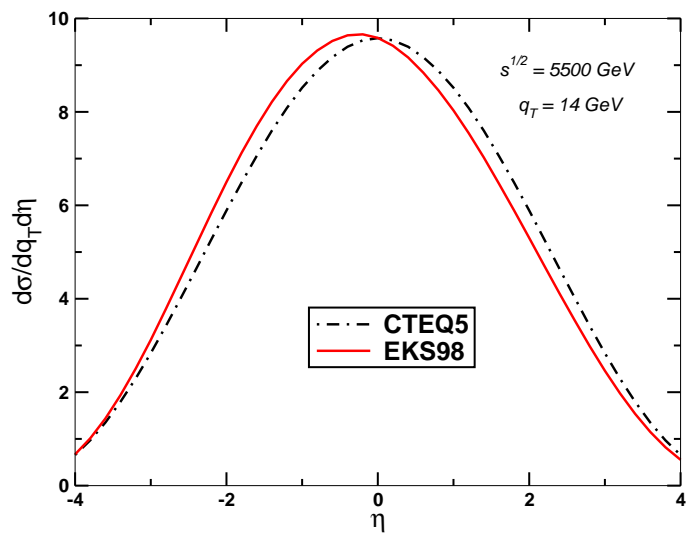}}
\end{tabular}}
\caption{The same as Fig \ref{fig0} for LHC ($\sqrt{s}=5500$ GeV).}
\label{fig1}
\end{figure}

At LHC energy ($\sqrt{s}=5.5$ TeV) and  p+Pb collisions,  the
cross section for the charm production probes the gluon
distribution in the region of $x \ge 3 \times 10^{-5}$ for $y \le
3$ \cite{pAhardprobes}. In this region, the EKS parameterization,
which is based on the DGLAP evolution equation and  global fits of
the DIS and Drell-Yan data above $Q^2 = 1$ GeV$^2$ and $x \ge
10^{-3}$, assumes that the nuclear gluon distribution behaves
similarly to the nucleon one, which implies that the ratio $R_g^A$
keeps constant. Consequently, it does not consider any new
dynamical effect associated to the high density of the medium in
this kinematic regime, which could modify $xg_A$ in comparison to
$xg_N$. This is a conservative assumption, since recent results
for forward rapidity indicate that the inclusion of the saturation
effects is necessary. However, as our goal is to provide a
baseline for future comparison, we use the EKS parameterization as
input in  our calculation. In Fig. \ref{fig1} we present our
predictions for the rapidity distributions for the $D$ meson
production at LHC energy. We have that at forward rapidities, the
suppression in the spectra decreases with increasing $q_T$, while
the anti-shadowing dominates at large negative values. The
crossover between the curves signalizes the rapidity value where
the nuclear shadowing begins to dominate and, with increasing
$q_T$, this point gets closer the central rapidity.

\begin{figure}
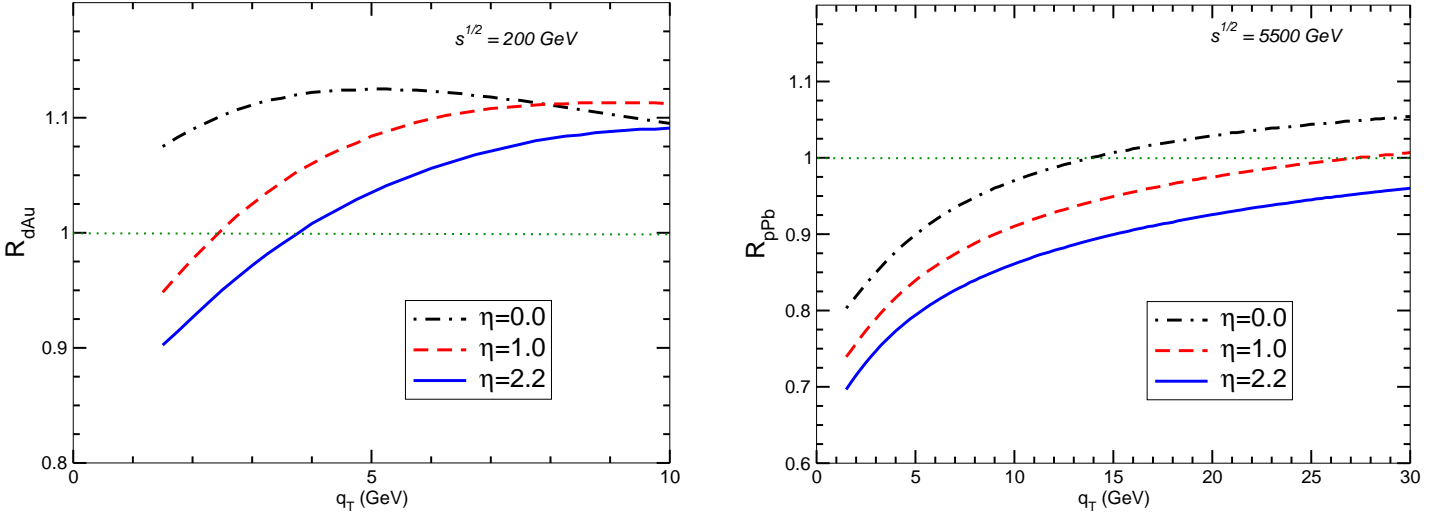

\centerline{
\begin{tabular}{ccc}
{\psfig{file=ratioheavy200.eps,width=90mm}} & &
{\psfig{file=ratioheavy5500.eps,width=90mm}}
\end{tabular}}
\caption{Nuclear modification ratio for D mesons in d+Au ($\sqrt{s}=200$ GeV) and p+Pb ($\sqrt{s}=5.5$ TeV) processes.}
\label{fig2}
\end{figure}

Our analysis regards to forward rapidities, since in this region the values
reached for $x_2$ becomes small enough to consider the saturation in the nuclear wave
function. The BRAHMS Collaboration has investigated the charged hadron production in
d+Au collisions at forward rapidities, with the values $\eta=1$, $\eta=2.2$ e $\eta=3.2$
\cite{brahms2}. Two of these values are considered to compute the evolution of the nuclear
modification ratio $R_{\mathrm{AB}}$, defined in Eq. (\ref{rabratio}),
in the transverse momentum. The prediction for mid-rapidity is also
shown for comparison. Even thought the NLO corrections
may affect the shape of the $q_T$ distributions, the higher-order corrections should largely
cancel out in this ratio. Our results are presented in Fig. \ref{fig2} for $\sqrt{s}=200$ GeV
and for $\sqrt{s}=5.5$ TeV, where to compute the denominator d+Au
processes were used in the first case and p+Pb processes in the later one.

At RHIC, as shown in the left panel of Fig. \ref{fig2}, the
spectra are enhanced in high-$q_T$ region, and the value of $q_T$
where the enhancement begins depends on the rapidity. If no
effects are present, we expect the ratio as unity. At
mid-rapidity, all the spectra is enhanced, weakening with
increasing $q_T$. With increasing rapidity, the spectra is
suppressed at low $q_T$, but becomes enhanced for higher values of
$q_T$, which is characteristic of the parameterization used. At
fixed rapidity, $x \propto m_T/\sqrt{s} \approx q_T/\sqrt{s}$ and
the values of $x$ increases with $q_T$, entering the
anti-shadowing region of the EKS parameterization. At very high
$q_T$, we expect that the ratio could fall below 1, due to EMC
effect in the EKS. At RHIC, this result suggests that in the
region where the validity of the perturbative treatment is
expected, $q_T>3$ GeV, the D meson spectra will be enhanced due to
nuclear anti-shadowing. This behavior is also present in charged
pion production in same energies \cite{vogtratio}. The 
preliminary open charm data from STAR Collaboration \cite{Ddatastar} at
mid-rapidity show this feature in the region 1 GeV $<q_T<$ 4 GeV.
In order to check our calculations,   we have calculated the  $D$ meson spectrum and verified that   
our results describe reasonably the experimental data \cite{Ddatastar}  in the region of interest for this study ($q_T
\ge 2$ GeV),  underestimating the data in the region of lower 
$q_T$, as expected, since we are not including an intrinsic transverse
momentum. 
This result is not shown since our main 
focus is the nuclear modification factor $R_{d\,Au}$,  which can be described using 
a leading order calculation. On the other hand, in the calculation  of 
the charmed hadron $q_T$ spectra,  the NLO corrections should be included, since it  modifies the shape of the  $q_T$ spectra as well as the normalization 
of the cross section. Both effects largely cancel in the calculation of 
the ratio $R_{d\,Au}$. It is important to emphasize that in the last Quark Matter Conference, the STAR collaboration has presented its preliminary results for the open charm spectrum in a broad transverse momentum region ($0 < q_T < 11$ GeV) \cite{Ddata_QM04}, with the measured open charm spectrum being much harder than the PYTHIA prediction,   which  is in line with the predictions from Ref. \cite{kharzeevNPA}. Such result  indicate that the saturation phenomenon may    be  important for the  heavy quark production at RHIC. However, more detailed studies related to the hadronization process are necessary before a definitive conclusion (See discussion in Ref. \cite{Ddata_QM04}). 

For LHC, as shown in the right panel of Fig. \ref{fig2}, the
behavior of $R_{p \, Pb}$ is similar for the three rapidities
analyzed. At low-$q_T$ the spectra is suppressed, and the exact
value where the enhancement takes place depends on the rapidity.
At mid-rapidity, this point is $q_T \approx 14$ GeV; for $\eta=1$,
it happens at $q_T \approx 27$ GeV, and this value increases for
$\eta=2.2$. Since the region to be studied at LHC is $q_T<14$ GeV,
this result suggests that a substantial suppression at positive
rapidities is due to nuclear shadowing effects.  Both panels in
Fig \ref{fig2} show a suppression in hadron spectra for $q_T<3$
GeV due to nuclear shadowing. However, the pQCD formulation might
not be valid anymore for a quantitative calculation in this small
$q_T$ region.

Finally, one expects that, because of their large mass,  radiative
energy loss for heavy quarks would be lower than for light quarks.
It occurs due to combined mass effects \cite{dokskhar,wang2zhang}:
the formation time of gluon radiation is reduced and their mass
also suppresses gluon radiation amplitude at angles smaller than
the ratio of the quark mass to its energy by destructive quantum
interference \cite{deadcone} - the dead-cone effect. Due to these
different energy losses, the ratio between hadrons with heavy
quarks and with light quarks can provide a tool to investigate the
medium formed in heavy ion collisions. The predicted consequence
of this distinct energy losses is an enhancement of this ratio at
moderately large transverse momentum, relative to that observed in
the absence of energy loss (A recent analysis for LHC energies is
given in Ref. \cite{daineseEPJC}).

\begin{figure}
\centerline{\psfig{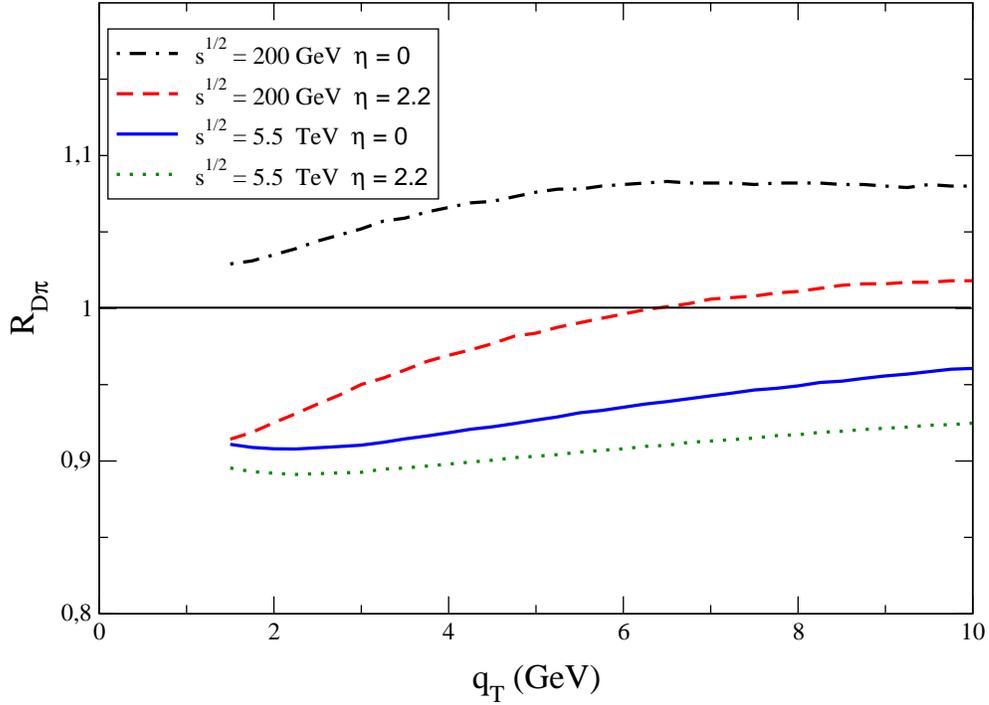}}
\caption{Ratio between D meson and pions, for RHIC and LHC energies and different rapidities.}
\label{fig3}
\end{figure}

In Fig. \ref{fig3}, results for the ratio between D mesons and $\pi$,
defined by
\vspace*{-0.6cm}
\begin{eqnarray}
{\mathcal{R}}_{D\pi}(\sqrt{s},q_T)= \frac{R_{\mathrm{hA}}^D}{R_{\mathrm{hA}}^\pi},
\end{eqnarray}
are presented considering hadron-nucleus collisions,  where the
final state effects, as energy loss, are minimal. The behavior of
the ratio considering only the shadowing in the nuclear
wavefunctions is presented. The ratio $R_{hA}^{\pi}$, defined as
in Eq. (\ref{rabratio}), is calculated  following  Ref.
\cite{accardiPLB} without the intrinsic transverse momentum. At
mid-rapidity, for RHIC, the D meson production is more enhanced
comparative to pions. This behavior also happens at large $q_T$,
for $\eta=2.2$. For LHC energies, the D meson production is
smaller that the pions, even at mid-rapidity. It suggests that the
shadowing predicted in Fig. \ref{fig2} at LHC for D mesons in p+Pb
processes is stronger that the shadowing for pion production, due
to the quadratic dependence on the gluon distribution present in
the charm production.

As a summary, the charmed meson production is studied using  a
perturbative approach. The rapidity distributions for d+Au
processes at RHIC and for p+Pb processes at LHC are computed, and
the disappearance of the asymmetry observed at low values of
transverse momentum was found for increasing $q_T$, since the
$x_2$ values increases with it. So, at high-$q_T$ and RHIC
energies, the D meson spectra is enhanced at forward rapidities.
For LHC, the analysis predicts a suppression for positive
rapidities due to nuclear shadowing, in the region $q_T<14 $ GeV,
even at mid-rapidity. We also studied the different behavior of
charmed mesons and light hadrons in $hA$ processes, where a
minimal energy loss are expected. Stronger nuclear effects for
heavy quarks were found, which cause an enhancement for D mesons
at mid-rapidity in RHIC, and their suppression for LHC, relative
to pion production. This feature is based in a conservative
perturbative  approach,  which assumes the validity of the
collinear factorization and that the EKS parameterization is
reasonable model for the nuclear effects. Although several points
deserve more detailed studies, we believe that it  can be used as
a baseline for the CGC dynamics, expected to be present in this
kinematical regime, as well as for future studies of jet quenching
effects in AA collisions.

\vspace*{0.4cm}
{\em Ackowledgements.} L.F.M. is very grateful to M. A. Betemps for
the discussions on the subject. This work was partially financed by CNPq and
FAPERGS, Brazil.

\newpage

\end{document}